\documentclass[aps,prb,showpacs,twocolumn]{revtex4}

\usepackage{amsmath}
\usepackage{amssymb}
\usepackage{graphicx}
\usepackage{textcomp}
\usepackage{dcolumn}
\usepackage{epstopdf}
\newcommand{\be}{\begin{equation}}
\newcommand{\ee}{\end{equation}}
\newcommand{\cms}{Co$_2$MnSi }

\begin{document}

\title{Thermodynamics of the Heusler alloy Co$_{2-x}$Mn$_{1+x}$Si: a combined density functional theory and cluster expansion study}

\author{Bj{\"o}rn H{\"u}lsen}
\affiliation{Fritz-Haber-Institut der Max-Planck-Gesellschaft, Faradayweg 4-6, D-14195 Berlin, Germany}

\author{Peter Kratzer}
\email[]{Peter.Kratzer@uni-duisburg-essen.de}
\affiliation{Fachbereich Physik, Universit{\"a}t Duisburg-Essen, Lotharstr. 1, D-47048 Duisburg, Germany}

\author{Matthias Scheffler}
\affiliation{Fritz-Haber-Institut der Max-Planck-Gesellschaft, Faradayweg 4-6, D-14195 Berlin, Germany}

\date{\today}

\begin{abstract}
Previous studies indicated that intrinsic point  defects 
play a crucial role for the density of states of ferromagnetic half-metals 
in the band gap region: 
at large concentrations, defect-derived bands might close the gap at the Fermi energy in the minority spin channel. In this work, structural disorder in the Co- and Mn-sublattices of the full Heusler alloy Co$_{2-x}$Mn$_{1+x}$Si ($-1 \le x \le 2$) is investigated with a cluster expansion approach, parametrized using all-electron density functional theory calculations. By establishing two separate cluster expansions, one for the formation energy and one for the total spin moment, we are in a position to determine the stability of different configurations, to predict new (also half-metallic) ground states and to extend the known Slater-Pauling rule for ideally stoichiometric  Heusler alloys to non-stoichiometric, Mn-rich compositions. This enables us to identify potentially half-metallic structures in the Mn-rich region. With the help of Monte Carlo simulations based on the cluster expansion, we establish theoretically that Co$_{2-x}$Mn$_{1+x}$Si close to the stoichiometric composition ought to show a high degree of structural order in thermodynamic equilibrium. Hence, samples prepared with the correct stoichiometry should indeed be half-metallic after thermal annealing. Moreover, we predict that adding a small amount of Mn to stoichiometric Co$_2$MnSi allows suppression of the thermally activated formation of detrimental Co antisites. At Mn-rich compositions ($x>1$), the ordered ground state structures predicted for zero temperature are found to be thermally unstable and to decompose into \cms and Mn$_3$Si above room temperature.
\end{abstract}

\maketitle

\section{Introduction}

Since their prediction in 1983 (\onlinecite{deGroot1983}) half-metallic ferromagnets have attracted much interest from a fundamental scientific point of view and as materials for spintronics devices. But there has been and still is a struggle to verify the existence of the gap in the minority spin channel experimentally. 
Following one line of research, the difficulty to observe a gap at finite temperatures is an intrinsic property of a ferromagnetic half-metal: In a many-particle description, thermal excitations of the electronic system formed by superpositions of majority spin states and virtual magnons appear in the Kohn-Sham band gap. These states filling up the half-metallic gap are referred to as
non-quasiparticle states\cite{Chioncel2003, Chioncel2006}.
Other researches have invoked extrinsic factors, such as structural defects, including atomic disorder, surface or interface states, to explain the low experimental spin polarization. \cite{Ritchie2003, Singh2004b, Kaemmerer2004, Singh2006} For half-metallic Heusler alloys first-principles calculations for bulk systems have shown that disorder in the sublattices\cite{Orgassa1999} or impurity bands induced by a high concentration of point defects\cite{Picozzi2004a} may close the gap. 

Here, we are interested in the full Heusler alloy \cms because it is a candidate for spintronics devices working at room temperature due to its large (predicted) Kohn-Sham gap,\cite{Ishida1995b, Picozzi2002} and its high measured Curie temperature $T_{\text{C}} = 985$~K. \cite{Webster1971, Brown2000} On the other hand, experimental values for the spin polarization of only $P=50-60\%$ (instead of the expected 100\%) have been measured for crystalline bulk and film samples using point-contact Andreev reflexion. \cite{Ritchie2003, Singh2004b,Singh2006}   
Somewhat higher values in the range of $P=60-90\%$ have been 
inferred from the tunneling magneto resistance ratio measured in tunnel junctions with a Co$_2$MnSi electrode\cite{Kaemmerer2004,Schmalhorst2005,Sakuraba2005,Sakuraba2006a,Ishikawa2006} at low temperatures, however, with a significant drop of $P$ around room temperature.\cite{Schmalhorst2007,Tsunegi2008}
Already in the early works, insufficiently ordered samples, in particular due to Co--Mn swaps, have been made responsible for the low values of $P$. Neutron diffraction measurements show that 14\% of the Mn sites are occupied with Co atoms in some samples. \cite{Ravel2002a} Density functional theory calculations using supercell geometries for \cms and Co$_2$MnGe show that some defects may change the electronic structure drastically,\cite{Picozzi2004a, Picozzi2007} but only four types of defects have been investigated so far.

The aim of this work is to study realistic, partially disordered \cms with {\it ab initio} accuracy over a wide concentration range and at elevated temperatures. We are especially interested in non-stoichiometric, Mn-rich \cms because the results of single defect calculations suggest that additional Mn atoms at Co sites may help to preserve the band gap. \cite{Picozzi2004a} Moreover, disorder, ``frozen in'' during sample preparation, could help to rationalize the non-ideal behavior of \cms observed in many experiments. In this study, we aim at identifying a range of compositions and annealing temperatures where half-metallicity can be expected. 

This paper is organized as follows: First, we briefly describe the cluster expansion method based on density functional theory calculations. It enables us to explore formation energies and magnetism in a huge configuration space. In Section 3, the properties of isolated defects and of the ground-state structures at zero temperature are reported. Moreover, we propose an extension of the Slater-Pauling rule that is useful in assessing the magnetic properties of Mn-rich Co$_2$MnSi. Finally, the cluster expansion of the formation energy is employed to perform Monte Carlo simulations both for nearly ideal \cms at finite temperatures, and for a newly discovered ordered alloy of composition Co$_2$Mn$_4$Si$_2$. In Section 4, we summarize our results and conclude.

\hspace*{1cm}

\section{Methods}

\subsection{The cluster expansion}

The cluster expansion (CE) method can be used to parametrize any physical property $F$ of a multicomponent system that depends uniquely on the atomic configuration. \cite{Sanchez1984} An excellent review of the method and its applications is given in Ref.~\onlinecite{Mueller2003}. Here, we only give a brief introduction of the underlying principle.

Let us consider a binary alloy A$_{1-x}$B$_x$ with $N$ lattice sites. Each site is described by an occupation variable $\sigma$ that is $+1 (-1)$ if a lattice point is occupied by atom A (B). A configuration of the whole crystal is then characterized by the occupation vector $\boldsymbol{\sigma} = \{ \sigma_1, \sigma_2, \ldots, \sigma_N \}$. The property $F$ can be expanded into a series with a structure similar to the Hamiltonian of the Ising model
\be
F(\boldsymbol{\sigma}) = J_0 + \sum_{i} J_i \sigma_i + \sum_{i,j} J_{ij} \sigma_i \sigma_j + \sum_{i,j,k} J_{ijk} \sigma_i \sigma_j \sigma_k + \ldots
\label{eq_basic_ce}
\ee
where the pairs, triplets and higher order terms are the so-called ``figures'' of the CE and the $J_i$ are the effective cluster interactions (ECI). Equation (\ref{eq_basic_ce}) is in principle exact \cite{Sanchez1984} but has to be truncated for practical reasons. The unknown parameters $J_i$ of the CE are determined from the results of a relatively small number of configurations ($\approx 50$) obtained through first-principles computations using a procedure that minimizes the mean-square deviations between the predicted and the calculated values of $F(\boldsymbol{\sigma})$ on subsets of the structures that have been calculated within DFT. \cite{vandeWalle2002a} This approach is also called the structure inversion method. \cite{Connolly1983} There remains the crucial question of how to identify the physically relevant interactions. Minimizing the least-squares error of the fit is not sufficient, since it does not guarantee the transferability of the fit to other (unknown) configurations. A more suitable measure is the cross-validation score (CVS). Here, we use the leave-one-out cross-validation score, defined by 
\be
C_{CVS}^2 = \frac{1}{n} \sum_i^n \bigl( F_i^{FP} - F_{(i)}^{CE} \bigr)^2,
\label{eq_cvs}
\ee
where the $F_{(i)}^{CE}$ are predicted by a least-squares fit to $n-1$ first-principles (FP) values \emph{excluding} $F_i^{FP}$. Therefore, $C_{CVS}$ gives an estimate of the predictive power of the fit. The set of figures that minimizes the cross-validation score is termed the optimal cluster expansion.

The method can be extended to multi-component systems. \cite{Sanchez1984} The CE Hamiltonian is most widely used for the calculation of alloy formation energies and as input for Monte Carlo simulations of thermodynamic quantities and phase diagrams, e.g. Cu-Au intermetallics \cite{Ozolins1998} or concentration of vacancies in the Al$_{1-x}$Li$_x$ alloy. \cite{vanderVen2005} The method has also successfully been applied for the calculation of magnetic properties of Fe-Co alloys \cite{Ortiz2006} and the computation of Curie temperatures of Mn-doped GaAs. \cite{Franceschetti2006}

\subsection{Computational details}

Density functional theory (DFT), using the highly accurate all-electron full-potential linearized augmented plane wave (FP-LAPW) method implemented in the Wien2k code \cite{Schwarz2003}, has been used to calculate total energies and magnetic spin moments of different compositions of Co$_x$Mn$_y$Si$_z$. The generalized gradient approximation (PBE96 \cite{Perdew1996}) to the exchange-correlation functional has been adopted because the GGA gives better results than the LDA (regarding lattice constants and bulk moduli) for Co$_2$MnSi according to Refs.~\onlinecite{Picozzi2002} and \onlinecite{Hashemifar2005}. For investigations of atomic defects, $2 \times 2 \times 2$ fcc supercells with 32 atoms and a  $6 \times 6 \times 6$ {\bf k}-point mesh have been used. As input for the cluster expansion, unit cells containing 4 to 32 atoms have been considered with dense {\bf k}-point meshes ($12 \times 12 \times 12$ for 4 atom cells, $6 \times 6 \times 6$ for 32 atom cells) to ensure convergence. The differences caused by the different {\bf k}-point meshes in different unit cell shapes are very small (e.g. the total energy difference between \cms in a 4-atom fcc unit cell and a 16-atom simple cubic unit cell is less than 0.1~meV/atom) and do not influence the results. The muffin-tin radii were set to $R_{MT}\text{(Co)} = R_{MT}\text{(Mn)} = 2.1$~bohr and $R_{MT}\text{(Si)} = 2.0$~bohr. 
The spin magnetic moments per atom quoted below have been calculated by integrating the spin densities obtained from converged  self-consistent calculations within the muffin-tin spheres, using these radii. 
Larger muffin-tin radii (touching spheres are achieved with $R_{MT}\text{(Co)} = R_{MT}\text{(Mn)} = R_{MT}\text{(Si)} = 2.3$~bohr) have been tested as well, but are unsuitable for the relaxation of the structures while they have negligible effect (less than 3\%) on the analysis of the atom-resolved spin moments. In the interstitial region a plane wave expansion cutoff of 15.5~Ry was used. All internal atomic coordinates were relaxed until the forces on the nuclei were smaller than 2~mRy/bohr.

\section{Results}

\subsection{Ideal \cms and the role of defects}

\begin{figure}
\includegraphics[angle=270,width=0.25\textwidth]{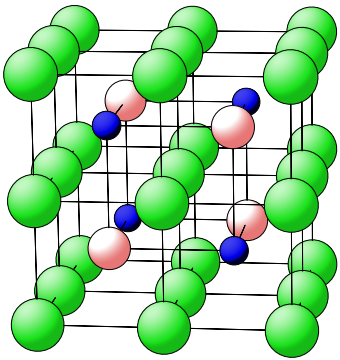}
\caption{\label{fig_cms_structure} (color online) The L2$_1$ structure of Co$_2$MnSi. Green (large) - Co atoms, pink (medium) - Mn atoms and blue (small) - Si atoms.}
\end{figure}

The L2$_1$ structure (common for all full Heusler alloys) of \cms is shown in Fig. \ref{fig_cms_structure}. Within density functional theory 
(both LDA and GGA) the half-metallicity of ideal \cms is well established and a summary of some calculated properties is given in Table \ref{tab_cms_properties}. Throughout the paper we use the term ``spin gap'' for the energy gap between the highest occupied state in the minority spin channel and the Fermi energy and the term ``band gap'' for the energy gap between the highest occupied state and the lowest unoccupied state in the minority spin channel.

A scheme for the hybridization of the $d$-bands in the full Heusler alloys that leads to the formation of the minority band gap has been established by Galanakis {\it et al.} (Ref. \onlinecite{Galanakis2002b}). Most of the Heuslers follow a Slater-Pauling (SP) rule that relates the total spin moment $M_{tot}$ of a compound to its number of valence electrons $N_V$ as follows:
\be
M_{tot} = N_V - 24 \,.
\label{eq_slater-pauling}
\ee
In other words, the SP rule states that the number of electrons in the minority spin channel is always 12 since this corresponds to a filling of four $sp$ bands, two $t_{2g}$ bands (occupied by three electrons each) and one $e_g$ band (occupied by two electrons) in the spin-down channel. These filled bands are separated from the empty bands by a hybridization gap. The difference in the number of valence electrons due to introducing other elements is accommodated in the majority spin channel only. A necessary (but not sufficient) condition for half-metallicity is an integer total spin moment per unit cell. The SP rule formulates a stricter, material specific (but still not sufficient) condition for the half-metallicity of Heusler alloys.
\begin{table}
\begin{ruledtabular}
\begin{tabular}{lD{.}{.}{1.3}D{.}{.}{1.3}D{.}{.}{1.3}D{.}{.}{1.3}}
 & \multicolumn{1}{c}{this work} & \multicolumn{1}{c}{Ref.~\onlinecite{Picozzi2002}} & \multicolumn{1}{c}{Ref.~\onlinecite{Galanakis2002b}} & \multicolumn{1}{c}{Ref.~\onlinecite{Ishida1995b}} \\
functional & \multicolumn{1}{c}{GGA} & \multicolumn{1}{c}{GGA} & \multicolumn{1}{c}{LDA} & \multicolumn{1}{c}{LDA} \\ 
\hline
\vspace{-2mm}
\\
$a$~(\AA)               & 5.629 & 5.634 & \multicolumn{1}{c}{exp.}  & 5.607 \\
$E_S$~(eV)              & 0.35  & 0.33  & \multicolumn{1}{c}{n/a}   & 0.36 \\
$E_G$~(eV)              & 0.84  & 0.81  & \multicolumn{1}{c}{n/a}   & 0.42 \\
$m_{\text{Co}}~(\mu_B$) & 1.07  & 1.06  & 1.02                      & \multicolumn{1}{c}{n/a}  \\
$m_{\text{Mn}}~(\mu_B$) & 2.91  & 2.92  & 2.97                      & \multicolumn{1}{c}{n/a}  \\
$m_{\text{Si}}~(\mu_B$) & -0.04 & -0.04 & -0.07                     & \multicolumn{1}{c}{n/a}  \\
$M_{tot}~(\mu_B$)       & 5.00  & 5.00  & 5.00                      & \multicolumn{1}{c}{n/a}  \\
\end{tabular}
\end{ruledtabular}
\caption{\label{tab_cms_properties}Calculated (DFT) lattice constant $a$, Kohn-Sham spin gap $E_S$, Kohn-Sham band gap $E_G$, atomic spin moments $m_i$ ($i$ = Co, Mn and Si) and total spin moment per unit cell $M_{tot}$ of bulk Co$_2$MnSi.}
\end{table}

Next we report results for several kinds of defects. The defect formation energy is calculated as \cite{CvandeWalle2004}
\begin{align}
E_f (X) = E_{tot}(X) - E_{tot}(\text{Co$_2$MnSi}) - \sum_i n_i \mu_i
\label{eq_defect_formation_energy}
\end{align}
where $E_{tot}(X)$ is the DFT total energy of the supercell containing defect $X$ and $E_{tot}(\text{Co$_2$MnSi})$ is the DFT total energy of ideal \cms in an equivalent supercell. The difference in the number of atoms to the stoichiometric composition is taken into account by $n_i$ that is $+1(-1)$ for an excess (deficiency) of atom of species $i$, and $\mu_i$ is the chemical potential of the reservoir (total energy per unit cell). We calculate the formation energies with respect to two different sets of reservoirs. The set A of reservoirs is the ``conventional'' one where we use the elements Co, Mn or Si in their most stable bulk form, i.e. ferromagnetic hcp Co, $\alpha$-Mn (calculated as antiferromagnetic $\gamma$-Mn plus a correction for the energy difference between $\alpha$- and $\gamma$-Mn taken from Ref.~\onlinecite{Hobbs2001}) and diamond Si. Since we are especially interested in the Co--Mn interactions in a Co- and Mn-rich regime, no elemental Si will be present, but compounds such as Mn$_3$Si or Co$_3$Si may be formed instead. Therefore,  we define a new set of reservoirs B where we eliminate $\mu_{\text{Si}}$ from Eq. (\ref{eq_defect_formation_energy}) using the laws of mass action for formation of Mn$_3$Si and Co$_3$Si. What enters in Eq. (\ref{eq_defect_formation_energy}) are now the total energies of hcp Co, $\alpha$-Mn, Co$_3$Si and Mn$_3$Si (all in their bulk form). The formation energies with respect to both sets of reservoirs, magnetic moments and spin gaps are shown in Table \ref{tab_defect_energies}. For the cases that have already been studied in Ref.~\onlinecite{Picozzi2004a} we find excellent agreement for the magnetic and electronic properties.

For most defective supercells the magnetic moments of atoms at regular sites are barely changed. The two exceptions are Si atoms at Co sites and the vacancies. Si atoms at Co sites increase the spin moments of the Co atoms surrounding them to $1.2~\mu_B$. In the case of the Si$_{\text{Co}}$ antisite this leads to a total spin moment of $41~\mu_B$ instead of the expected $39~\mu_B$. In case of a vacancy, the moments of the neighboring atoms are reduced, both for Co ($0.8 - 0.9~\mu_B$) and Mn ($2.7 - 2.8~\mu_B$) neighbors. A similar behavior is observed for Si, Mn and Co vacancies. The defects themselves show different behavior. Co atoms at Mn sites have magnetic moments of $1~\mu_B$ (Co--Mn swap) and $1.2~\mu_B$ ($\text{Co}_{\text{Mn}}$ antisite). Co atoms at Si sites show an increased magnetic moment of about $1.6~\mu_B$ (for swap and antisite). In agreement with earlier calculations\cite{Ozdogan2007a}, we find that Mn atoms at Co sites (as part of a swap and as antisite) couple antiparallel to atoms at regular sites with a magnetic moment of around $-1~\mu_B$ which leads to reduced total magnetic moments compared to the ideal case. 

For a $\text{Mn}_{\text{Co}}$ antisite in a 32-atom supercell we have done a calculation with a fixed total spin moment (FSM) 
of the supercell of $42~\mu_B$ which corresponds to a substitution of $m_{\text{Co}} \approx 1~\mu_B$ with $m_{\text{Mn}} \approx 3~\mu_B$. In this case the Mn antisite atom changes its magnetic moment from $-1~\mu_B$ to $2.3~\mu_B$ and the Co atoms (at regular sites) show slightly increased spin moments of about $1.15~\mu_B$. The energy difference between the FSM calculation and the free one with $M_{tot} = 38~\mu_B$ is about 1.1~eV. This reflects that the antiparallel coupling of the Mn atom at a Co site is strongly preferred over parallel coupling.

\begin{figure}
\includegraphics[width=\columnwidth,angle=0]{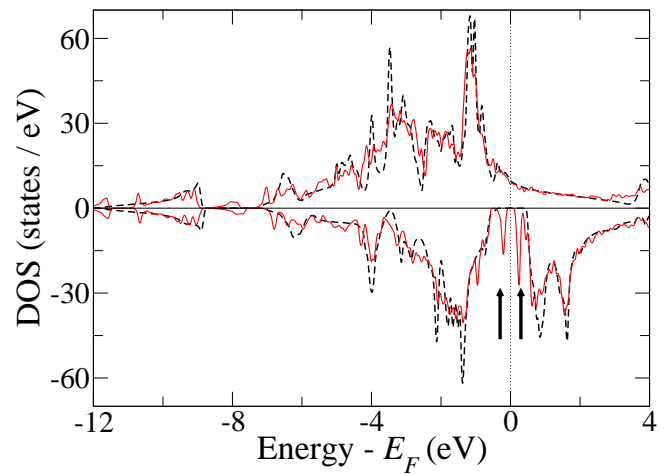}
\caption{\label{fig_co-si_swap_dos} (color online) The total density of states of ideal \cms (black dotted line) and the Co--Si swap (red line). The crucial difference lies in the states in the band gap , i.e. between $-0.5$eV and $+0.35$eV,  indicated by the two arrows, that are derived solely from the Co atoms at Si sites.}
\end{figure}

As an example for the impact on the electronic structure the total density of states (DOS) of the Co--Si swap is compared to the ideal case in Fig. \ref{fig_co-si_swap_dos}. There are two defect states close to the Fermi energy that are solely derived from the Co atom at Si site. One of them is occupied and lies slightly below $E_F$, the other one is unoccupied. Due to these states the remaining spin gap is very small. The DOS of the $\text{Co}_{\text{Si}}$ antisite is similar to that of the Co--Si swap showing also the two defect states. But in this case the occupied state lies at the Fermi energy and closes the spin gap.

We note that the described position of levels with respect to the Fermi energy refers to the Kohn-Sham energy levels calculated within the DFT-GGA approximation. We expect the qualitative picture, e.g. the orbital character of the defect states, to remain correct if methods beyond DFT-GGA were used. The relative energy position of Mn and Co orbitals is correctly described by DFT-GGA, and hence the predicted Co orbital character of highest occupied spin-down states will persist even in a higher-level method. However, the exact position of the Fermi level relative to the defect-derived states may depend on the treatment of electronic exchange and correlation effects. For example, in recent calculations performed with the GGA$+U$ method it was observed that the position of the Fermi level can be shifted relative to the band edges in the spin down channel if the value of $U$ is varied. \cite{Kandpal2006a}

Taking into account our results together with the known results for Co--Mn swaps, Co$_{\text{Mn}}$, Mn$_{\text{Co}}$ antisites\cite{Picozzi2004a}, Mn$_{\text{Si}}$ and Si$_{\text{Mn}}$ antisites\cite{Galanakis2006a} and vacancies\cite{Ozdogan2007} (all obtained in the framework of DFT) we can conclude that defects with Co atoms at Mn or Si sites may close the spin gap while all other defects have only minor effects on the electronic structure and are harmless for the half-metallicity.
However, the vacancies lead to a considerable reduction in the spin gap (see Table~\ref{tab_defect_energies}), and have been predicted\cite{Ozdogan2007} to disturb the half-metallicity for concentrations $\ge 10$\%.
\begin{table}
\begin{ruledtabular}
\begin{tabular}{lD{.}{.}{2.2}D{.}{.}{2.2}D{.}{.}{2.2}D{.}{.}{2.2}}
 & \multicolumn{1}{c}{$E_f^A$~(eV)} & \multicolumn{1}{c}{$E_f^B$~(eV)} & \multicolumn{1}{c}{$M_{tot}~(\mu_B)$} & \multicolumn{1}{c}{$E_S$~(eV)} \\
 \hline
$\text{Co}_{\text{Mn}}$ antisite & 0.75 & 0.65 & 38.0 & 0.05 \\
$\text{Co}_{\text{Si}}$ antisite & 2.25 & 1.26 & 41.0 & 0.00 \\
$\text{Mn}_{\text{Co}}$ antisite & 0.36 & 0.45 & 38.0 & 0.39 \\
$\text{Mn}_{\text{Si}}$ antisite & 1.69 & 0.40 & 43.0 & 0.42 \\
$\text{Si}_{\text{Co}}$ antisite & 1.78 & 2.77 & 41.0 & 0.59 \\
$\text{Si}_{\text{Mn}}$ antisite & -0.21& 1.08 & 37.0 & 0.33 \\
\hline
Co--Mn swap  & 1.20 & 1.20 & 36.0 & 0.20 \\
Co--Si swap  & 3.84 & 3.84 & 42.0 & 0.12 \\
Mn--Si swap  & 1.17 & 1.17 & 40.0 & 0.43 \\
\hline
$V_{\text{Co}}$ vacancy  & 1.04 & 1.04 & 37.0 & 0.26 \\
$V_{\text{Mn}}$ vacancy  & 1.43 & 1.43 & 33.0 & 0.15 \\
$V_{\text{Si}}$ vacancy  & 3.74 & 2.72 & 36.0 & 0.14 \\
\end{tabular}
\end{ruledtabular}
\caption{\label{tab_defect_energies}Formation energies $E_f$ according to eq. (\ref{eq_defect_formation_energy}) with respect to reservoirs A (Co, Mn and Si) and B (Co, Mn, Co$_3$Si and Mn$_3$Si), total spin moments $M_{tot}$ and Kohn-Sham spin gap $E_S$ of some possible defects in supercells with 32 atoms. Stoichiometric \cms in a 32-atom supercell has a total spin moment of $40~\mu_B$.}
\end{table}

\subsection{CE for pseudo-binary Co$_{2-x}$Mn$_{1+x}$Si}

The results for single atomic defects show that the electronic and magnetic properties of \cms are strongly altered by defects involving Co (leading to defect states close to or even at the Fermi energy) and Mn (magnetic moment antiparallel to regular Co and Mn) atoms. 
We verified by explicit calculations that other defects, such as vacancies, Mn--Si swaps or Si antisites have less impact on the electronic structure (i.e. they do not induce defect states in the half-metallic gap) and on the magnetic properties (no antiparallel coupling). Since the formation energies of a Co--Si swap and a $\text{Co}_{\text{Si}}$ antisite are much larger than the corresponding Co--Mn swap and $\text{Co}_{\text{Mn}}$ antisite (for both sets of reservoirs considered in Tab. \ref{tab_defect_energies}), the former two are neglected in the following. Thus, we regard the material as a pseudo-binary alloy where only the interactions between the Co and Mn atoms are taken into account while the Si sublattice is kept fixed.

The cluster expansions are constructed using the ATAT software. \cite{vandeWalle2002b, vandeWalle2002c} This `alloy theoretic automated toolkit' generates structures on the given parent lattice and uses the DFT values (e.g. total energy) of these structures to fit all interactions up to a given maximum number of points of the CE figures (here six). It compares the cross-validation scores of the different CEs and chooses the CE with the lowest one.

\begin{figure}
\includegraphics[width=\columnwidth]{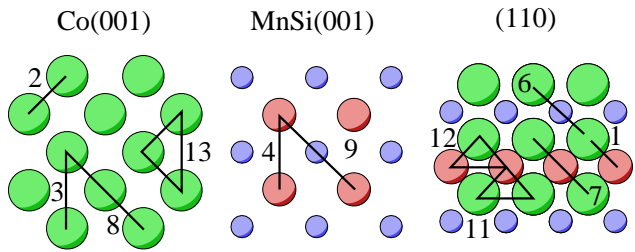}
\caption{(color online) Optimal CE figures labeled by their index. The green (large) circles are Co sites, the pink (medium) circles  Mn sites, and the blue (small) circles Si sites. The triples 14 and 15 (not shown) are the smallest possible equilateral triangles in Co(111) and Mn(111) planes. Co-Mn pair 5 and Co-Co-Mn triple 10 lie outside the lattice planes shown in the figure.}
\label{fig_ce_figures}
\end{figure}

\begin{figure}
\includegraphics[width=0.7\columnwidth,angle=270]{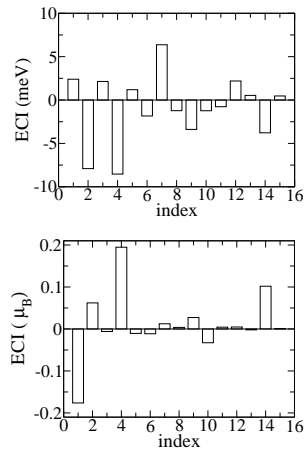}
\caption{Effective cluster interactions (ECI) for the CE for the formation energy (top), and for the total spin moment (bottom). For the indexing of the CE figures, cf. Fig.~\protect\ref{fig_ce_figures}.}
\label{fig_ce_values}
\end{figure}

In the following the CE for the formation energies of the compounds is called ECE and the one for the total spin moment is called MCE. The total energies and spin moments of 60 structures have been computed with DFT-GGA for the fit of the interactions. The optimal set of figures for both CEs consists of 9 pairs and 6 triplets  
(most of them shown in Fig. \ref{fig_ce_figures}) with cross validation scores $C_{CVS}^{ECE} = 5.8$~meV/atom and $C_{CVS}^{MCE} = 0.05~\mu_B/\text{atom}$. 
As will be seen later, $C_{CVS}^{ECE}$ is of the same order of magnitude as the ordering energies of Mn-rich structures. 
The strengths of the effective cluster interactions (ECI) are shown in Fig.~\ref{fig_ce_values} both for the CE of the total energy and the CE of the total spin moment.   In the energy expansion, the leading attractive interactions (negative values) are the pairs 2 and 4 and the triple 14. These interactions ensure that the sites in the Co(100) plane, the Mn sites in the MnSi(100) plane, and the sites in the Co(111) plane, respectively, are preferentially occupied by like species of the suitable kind. 
Tests with a CE expansion containing more long-ranged pair interactions, or additional triples or quadruples, didn't lead to a significant reduction of $C_{CVS}^{ECE}$.

One has to keep in mind that Co$_3$Si and Mn$_3$Si are the reference structures and reservoirs for the ECE and that the Si sublattice is fixed. Therefore, the general eq. (\ref{eq_defect_formation_energy}) is adopted for this special case, by eliminating $\mu_{\text{Si}}$ in the way described above. Every structure contains exactly 25\% Si and the chemical potentials are now the total energies per unit cell of Co$_3$Si and Mn$_3$Si. The formation energy per atom of a Co$_{2-x}$Mn$_{1+x}$Si compound is then defined as
\begin{align}
E^{ECE}_f(\text{Co}_a\text{Mn}_b\text{Si}_{(a+b)/3}) = \qquad \qquad \qquad \qquad \nonumber \\
\frac{1}{a+b} \Bigl[ E(a,b) - \frac{a}{3}E(\text{Co}_3\text{Si}) - \frac{b}{3}E(\text{Mn}_3\text{Si}) \Bigr]
\label{eq_formation_CE}
\end{align}
where $E(a,b)$ is the total energy of the structure calculated with DFT.

\begin{figure}
\includegraphics[width=6.5cm,angle=0]{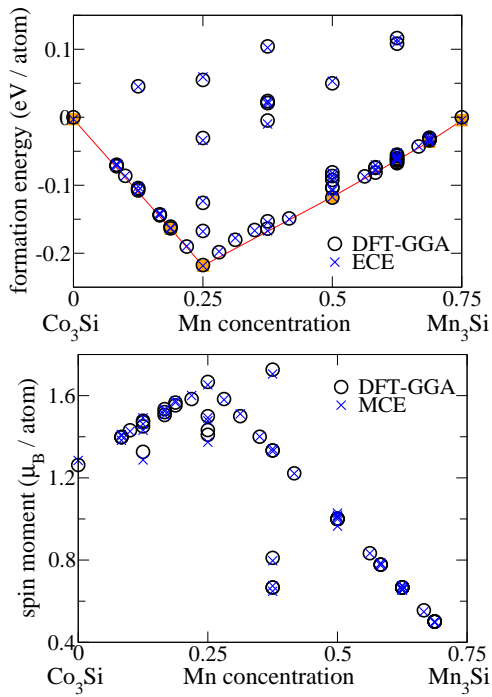}
\caption{\label{fig_dft_ce}Comparison of the values from DFT-GGA and from the CEs (top: formation energy, bottom: total spin moment). The excellent agreement between the the values demonstrates the quality of the two fits.}
\end{figure}

In Fig. \ref{fig_dft_ce}, the values from the DFT-GGA calculations are compared to the CE values. Excellent agreement is found. Additionally, a ground state search has been performed and the ground state line is included in the upper part of Fig. \ref{fig_dft_ce}. For three structures $\alpha$, $\gamma$ and $\beta$ with concentrations $x(\alpha) < x(\gamma) < x(\beta)$ that have the lowest formation energies at their respective concentrations structure $\gamma$ has to fulfill the condition
\be
E_f^{ECE}(\gamma) < \frac{x(\gamma) - x(\beta)}{x(\alpha) - x(\beta)} E_f^{ECE}(\alpha) + \frac{x(\gamma) - x(\alpha)}{x(\beta) - x(\alpha)} E_f^{ECE}(\beta)
\label{eq_gs_bedingung}
\ee
to be part of the ground state line at $x(\gamma)$. \cite{Mueller2003} In other words, the ground state line is the lower convex envelope of all data points in the formation energy diagram. With the ECE Hamiltonian the formation energies of all combinatorial possibilities in a unit cell with 32 atoms (24 Co/Mn sites, $2^{24}\approx 16.8$ million) and of $10^7$ structures in a unit cell with 128 atoms (96 Co/Mn sites) where the sites have been randomly occupied with Co or Mn have been computed. The well-known L2$_1$ structure of \cms is the most stable configuration. Four new ordered ground states at zero temperature are predicted at Mn concentrations of $x=0.1875$, $x=0.5$, $x=0.625$, and $x=0.6875$.  
At these concentrations, slight (almost invisible) cusps appear, deviating by about 5 mV/atom from the convex envelope. 
Since these deviations are very small, test calculations have been performed, demonstrating that the formation energies predicted by the ECE agree with the DFT-GGA values  (cf. Fig.~\ref{fig_dft_ce}).
For concentrations different from the ordered ground states, the lowest energy state of the system 
at zero temperature consists of two regions of coexisting phases. The fact that the data points in Fig.~\ref{fig_mce} do not reach down to the convex envelope at all concentrations is due to the finite size of the unit cell used, which doesn't allow for phase separation to occur.

\subsection{The ordered ground states}

The ordered ground state at the Co-rich side is a single $\text{Co}_{\text{Mn}}$ antisite in a 16-atom simple cubic unit cell (Co$_9$Mn$_3$Si$_4$). It has a broad Co-derived band crossing the Fermi energy, leading to a small spin polarization and rendering this structure uninteresting for potential spintronics applications.

\begin{figure}
\includegraphics[angle=270, width=0.4\textwidth]{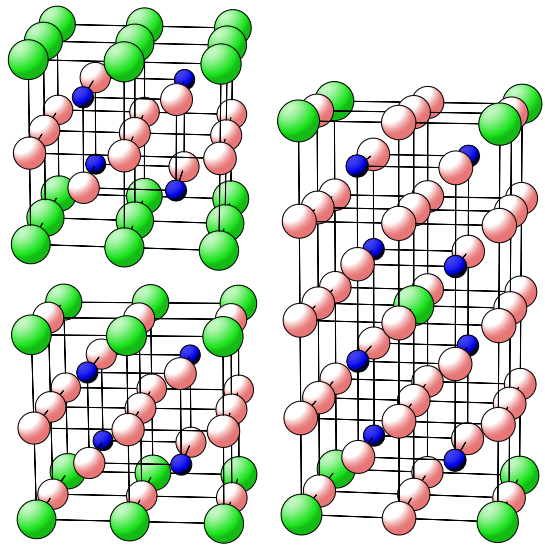}
\caption{\label{fig_new_gs} (color online) Structures of the Mn-rich ground states. top left: Co$_2$Mn$_4$Si$_2$, bottom left: Co$_2$Mn$_{10}$Si$_4$ and right: CoMn$_{11}$Si$_4$. Green (large) - Co atoms, pink (medium) - Mn atoms, and blue (small) - Si atoms.}
\end{figure}

We will concentrate on the ordered ground states at the Mn-rich side. They have highly symmetric structures that show a certain trend in the incorporation of the Mn atoms at Co sites as shown in Fig. \ref{fig_new_gs}. First, in Co$_2$Mn$_4$Si$_2$ each second layer of Co atoms is completely replaced by Mn atoms. Increasing the Mn concentration further leads to additional Mn rows in the remaining Co planes (Co$_2$Mn$_{10}$Si$_4$ and CoMn$_{11}$Si$_4$). These ground states follow the general trend in the physics of alloys that stable structures have relatively small and highly symmetric unit cells.

\begin{figure}
\includegraphics[width=\columnwidth,angle=0]{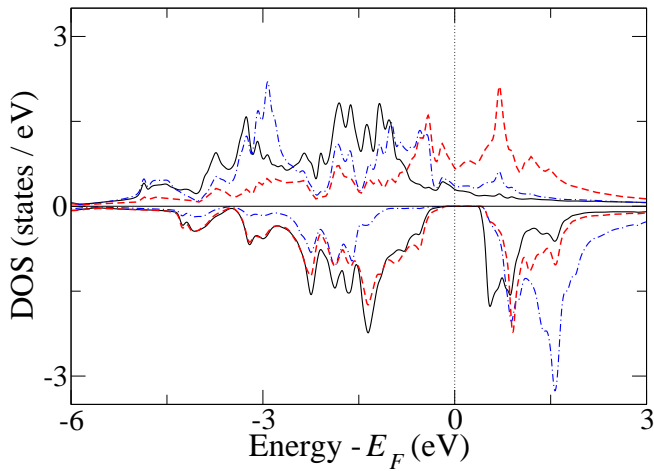}
\caption{\label{fig_37_dos} (color online) The density of states projected on the $d$-orbitals of the predicted half-metallic ground state Co$_2$Mn$_4$Si$_2$ (black line - Co, blue dash-dotted line - regular Mn, and red dashed line - Mn atom at Co site). In the minority spin channel the DOS of the Co atom and of the Mn atom at the Co site are almost identical.}
\end{figure}

Ideal \cms is a half metal according to DFT calculations. The electronic structures of the other ground states show one remarkable result. The total DOS of Co$_2$Mn$_4$Si$_2$ has a gap in the minority spin channel around $E_F$ and is therefore a half metal, too. Investigating the orbital-resolved DOS (shown  in Fig. \ref{fig_37_dos}) reveals that the spin-down DOS of the Mn atom at Co site is almost identical to the spin-down DOS of Co. This behavior is also reflected in the magnetic moments of the Mn atoms at Co sites. They align antiparallel to the regular Co and Mn magnetic moments, and their magnitude ranges from $-0.6~\mu_B$ to $-0.7~\mu_B$. The integration of the DOS from $-\infty$ to $E_F$ confirms that the missing electrons (due to replacing Co with Mn) in this structure are taken only from the majority spin channel. This complies with the Slater-Pauling rule for the full Heusler alloys mentioned earlier.

\subsection{Extended Slater-Pauling rule}

In the supercell calculations, $\text{Mn}_{\text{Co}}$ antisites reduce the spin moment by $2~\mu_B$ (from $+1~\mu_B$ to $-1~\mu_B$) which corresponds exactly to the reduction of the number of valence electrons (from 7 Co $d$ electrons to 5 Mn $d$ electrons). This suggests that the Slater-Pauling rule may still be valid even for non-stoichiometric, Mn-rich compositions Co$_{2-x}$Mn$_{1+x}$Si. We have performed DFT-GGA calculations of Mn-rich structures to verify the 
linear relation implied by the Slater-Pauling rule between the number of valence electrons and the total spin moment. These data are summarized in Table \ref{tab_ce_mn-rich}. The spin gap decreases with increasing Mn ratio until it is finally closed at a Mn concentration of about 60\%. Thus, there is a wide Mn concentration range from 25\% to 60\% where the half-metallicity is preserved, and the total spin moment per unit cell remains to be an integer multiple of $\mu_B$.
\begin{table}
\begin{ruledtabular}
\begin{tabular}{lrD{.}{.}{2.2}D{.}{.}{2.2}D{.}{.}{1.2}}
 structure & $N_V$ & \multicolumn{1}{c}{$M_{tot}~(\mu_B)$} & \multicolumn{1}{c}{$N_V - M_{tot}$} & \multicolumn{1}{c}{$E_S$~(eV)} \\
\hline
Co$_2$MnSi            & 29  & 5.0  & 24   & 0.35 \\        
Co$_{15}$Mn$_9$Si$_8$ & 230 & 38.0 & 192  & 0.38 \\
Co$_7$Mn$_5$Si$_4 $   & 114 & 18.0 & 96   & 0.35 \\
Co$_8$Mn$_7$Si$_5$    & 141 & 21.0 & 120  & 0.34 \\
Co$_6$Mn$_6$Si$_4$    & 112 & 16.0 & 96   & 0.20 \\
Co$_6$Mn$_6$Si$_4$    & 112 & 16.0 & 96   & 0.32 \\
Co$_4$Mn$_5$Si$_3$    & 83  & 11.0 & 72   & 0.19 \\
Co$_2$Mn$_4$Si$_2$    & 54  & 6.0  & 48   & 0.13 \\
Co$_2$Mn$_4$Si$_2$    & 54  & 6.0  & 48   & 0.14 \\
Co$_4$Mn$_8$Si$_4$    & 108 & 12.0 & 96   & 0.11 \\
Co$_3$Mn$_9$Si$_4$    & 106 & 10.0 & 96   & 0.08 \\
Co$_2$Mn$_7$Si$_3$    & 79  & 7.0  & 72   & 0.09 \\
Co$_2$Mn$_7$Si$_3$    & 79  & 7.0  & 72   & 0.07 \\
Co$_2$Mn$_7$Si$_3$    & 79  & 7.0  & 72   & 0.09 \\
Co$_2$Mn$_{10}$Si$_4$ & 104 & 8.0  & 96   & 0.01 \\
Co$_2$Mn$_{10}$Si$_4$ & 104 & 8.0  & 96   & 0.00 \\
Co$_2$Mn$_{10}$Si$_4$ & 104 & 8.0  & 96   & 0.04 \\
CoMn$_8$Si$_3$        & 77  & 5.0  & 72   & 0.00 \\
CoMn$_{11}$Si$_4$     & 102 & 6.02 & 95.98 & 0.00 \\
CoMn$_{11}$Si$_4$     & 102 & 6.02 & 95.98 & 0.00 \\
CoMn$_{11}$Si$_4$     & 102 & 6.01 & 95.99 & 0.00 \\
\end{tabular}
\caption{\label{tab_ce_mn-rich}Number of valence electrons $N_V$, total spin moment $M_{tot}$, and spin gap $E_S$ of some Mn-rich structures.}
\end{ruledtabular}
\end{table}

The relation between the number of valence electrons and the total spin moment can be expressed as an extended Slater-Pauling rule
\be
M_{tot} = N_V - 24 n_{\text{Si}}
\label{eq_slater-pauling_ext}
\ee
where $n_{\text{Si}}$ is the number of Si atoms in the unit cell. The Mn-rich structures show similar orbital-resolved DOS as in Fig. \ref{fig_37_dos}; Mn atoms at Co sites mimic the DOS of the Co atoms in the minority-spin channel. Therefore, the original hybridization scheme proposed in Ref.~\onlinecite{Galanakis2002b} is still valid and the SP rule also holds for the large unit cells of these compounds. The factor $n_{\text{Si}}$ takes the enlarged size of the unit cell into account.

\begin{figure}
\includegraphics[width=\columnwidth,angle=0]{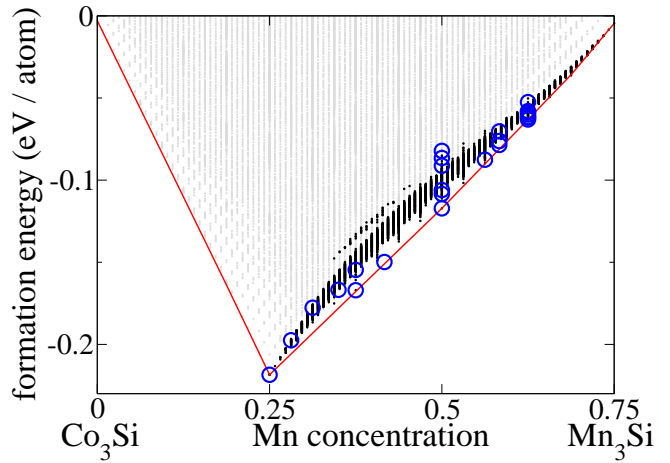}
\caption{\label{fig_mce} (color online) Direct enumeration of the formation energies according to eq. (\ref{eq_formation_CE}) of 27 million structures. The black dots belong to structures that obey the extended Slater-Pauling rule of eq. (\ref{eq_slater-pauling_ext}). For the structures marked by blue circles it has been confirmed by DFT-GGA calculations that they have a spin gap and and satisfy eq. (\ref{eq_slater-pauling_ext}).}
\end{figure}

The CE for the total spin moment allows us to check millions of unknown structures whether they fulfill the extended SP rule. The result is plotted in Fig. \ref{fig_mce} where the formation energies of structures that satisfy eq. (\ref{eq_slater-pauling_ext}) within a tolerance of $0.05~\mu_B / \text{atom}$ (given by the cross-validation score) are highlighted. A large region of potential half-metallic compositions is identified.

\subsection{Monte Carlo simulations}

From our defect calculations it has become clear that even ideally stoichiometric \cms shows a spin gap only if it is highly ordered. Hence the question arises how the properties of \cms are affected by thermal disorder at finite temperature. The ECE offers an excellent starting point to address this question by means of Monte Carlo simulations. The thermodynamic quantity governing the behavior of the material at finite temperature is the free energy, $F = E - TS$. While configurational entropy is treated implicitly in the Monte Carlo simulation, the contributions from vibrational or magnetic excitations should be included in the free energy functional employed in performing the MC simulations.
 
As a first example, we restrict ourselves to \cms with ideal stoichiometry. In this case, thermal disorder will manifest itself by atomic swaps where we restrict ourselves to Co--Mn swaps that are detrimental to the half-metallic property. For a Monte Carlo simulation of stoichiometric \cms at finite temperature, the crucial input data are the free energy differences associated with defect formation. Analyzing the role of vibrational and magnetic excitations, we will conclude that none of them contributes significantly to free energy {\em differences}.
Concerning the vibrational contributions, we argue that both the mass and the bond strength of Mn and Co atoms are very similar. Therefore, we expect only very small changes in the phonon 
density of states between ideal and defective samples. The phonon contribution to free energy  differences is therefore neglected.
As far as magnetic excitations are concerned, we note that most defects couple ferromagnetically to the (ferromagnetic) host material. Far below the Curie temperature of $T_{\text{C}} = 985$~K, the magnetic excitations will consist of small oscillations around the direction of the average magnetization, both for the ideal and the defected materials. Because of this similarity, the magnetic contributions to the free energy for both situations tend to cancel. The only defect whose magnetic moment aligns antiparallel to the host material is the Mn atom on a Co site. For this case, we calculated the energy to invert the Mn magnetic moment to be $E_{mag} = 1.1$~eV  (additional to the structural formation energy). Hence the Mn magnetic moment cannot flip freely, but, also in this case, will perform only small oscillations around the ground state. Therefore, we neglect the magnetic contributions to the free energy differences as well. 

\begin{figure}
\includegraphics[width=\columnwidth,angle=0]{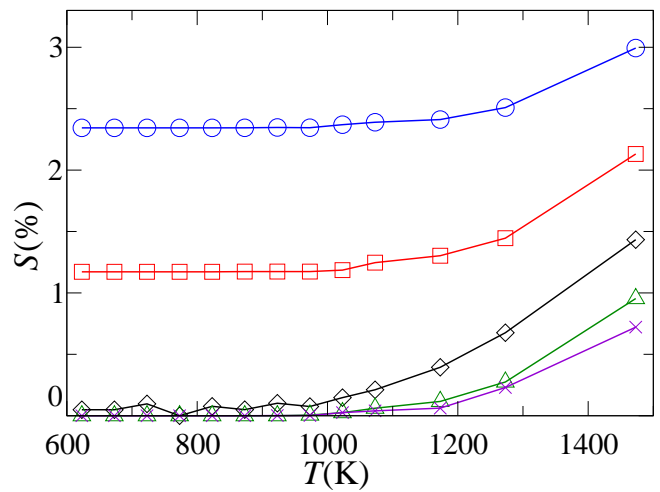}
\caption{\label{fig_lro_vs_t} (color online) The temperature dependence of the fraction of Co atoms at Mn sites $S$ for different compositions in the simulation cell (blue circles: 4144 Co and 2000 Mn atoms, red squares: 4120 Co and 2024 Mn atoms, black diamonds: 4096 Co and 2048 Mn atoms (stoichiometric composition), green triangles: 4072 Co and 2072 Mn atoms, violet crosses: 4048 Co and 2096 Mn atoms). At temperatures below 1000~K the stoichiometric sample shows only 1\textperthousand{} of Co defects and is almost perfectly ordered.
Even above 1000~K, the Co antisite formation may be suppressed by excess Mn atoms. }
\end{figure}

We perform Monte Carlo (MC) simulations of stoichiometric \cms to sample the configurational space in thermodynamic equilibrium. Starting from a random occupation of the lattice, the Mn atoms swap their sites with nearest- or next-nearest Co neighbors. The energies of the configurations are computed with the ECE Hamiltonian, and the new configuration is accepted according to the Metropolis criterion. The MC results presented here have been obtained in a simulation cell containing 6144 sites for Co and Mn atoms.
The simulations show that the concentration of Co--Mn swaps is very low, below 1 \textperthousand{} at temperatures of up to 1000~K (Fig.~ \ref{fig_lro_vs_t} , diamond symbols).  Tests with an increased cell size of 12,000 sites gave the same results. 
We note that the interaction energy between Mn atoms at Co sites is positive, but small (1.7~meV/atom from DFT-GGA, 2.9~meV/atom from ECE). At the low defect concentrations observed, the Co--Mn swaps can be treated as statistically independent and their concentration is given by
\be
c_{\text{Co-Mn}} = \frac{1}{\sqrt{2}} \exp \left( - \frac{E_f}{2 k_B T} \right).
\label{eq_pair_defects}
\ee
The comparison of the MC results with the analytical solution shows excellent agreement, confirming that it is indeed justified to neglect defect interactions in this case.

Next, MC simulations are performed for samples that are slightly non-stoichiometric. We define the relative fraction $S$ of the number of Co atoms $n_{\text{Co}}$ at Mn sites $N_{\text{Mn}}$, 
\be
S = \frac{n_{\text{Co}}}{N_{\text{Mn}}}.
\label{eq_lro}
\ee
Again, Fig. \ref{fig_lro_vs_t} shows that the system tends to an almost perfectly ordered state for temperatures below 1000~K. For higher temperatures, some additional Co antisites are created due to thermal disorder. For Mn-rich compositions, we note that the increase of thermally created Co defects is somewhat smaller than for the stoichiometric sample, i.e., the presence of excess Mn allows the suppression of Co defect formation. Moreover, we observe that the positive interaction energy  between Mn atoms at Co sites leads to clustering of Mn antisites.

\begin{figure}
\includegraphics[width=0.7\columnwidth,angle=270]{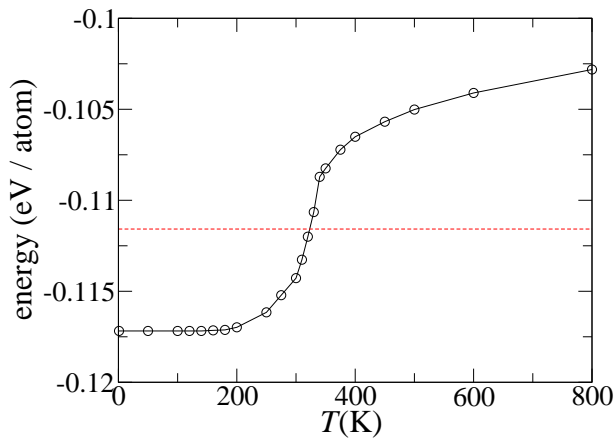}
\caption{\label{fig_37_ene_vs_T}Energy per atom vs. temperature from a MC simulation with 4000 Co and 8000 Mn atoms in the cell corresponding to ground state Co$_2$Mn$_4$Si$_2$ at $T = 0$~K. The dashed line shows the formation energy difference between the ground state Co$_2$Mn$_4$Si$_2$ and a linear interpolation between ideal \cms and Mn$_3$Si. The intersection point coincides with the maximum slope of $c_V = \partial  E / \partial T|_V$ that indicates the latent heat of the phase transition from the ordered ground state to phase separation at $T \approx 320$~K.}
\end{figure}

We also perform MC simulations for a sample whose composition matches the newly predicted  ground state Co$_2$Mn$_4$Si$_2$ in order to assess the thermal stability of this ordered structure. 
First, we note that the energy difference between the formation energy of ground state Co$_2$Mn$_4$Si$_2$ and a linear interpolation between the energies of ideal \cms and Mn$_3$Si is very small (5.6~meV/atom).  The MC simulation shows that this leads to a phase separation of Co$_2$Mn$_4$Si$_2$ into \cms and Mn$_3$Si in thermodynamic equilibrium for temperatures $T \geq 320$~K (Fig. \ref{fig_37_ene_vs_T}). Thus, the potentially half-metallic ordered structures that we predict in a wide composition range may not be practically relevant. Rather, the samples in this composition range could be described by precipitates of Mn$_3$Si (which is not a half metal) in a matrix of half-metallic Co$_2$MnSi. The electrical properties will then depend on the size and the concentration of these precipitates and on their percolation length.

\section{Conclusions}

In this work, cluster expansions on the basis of DFT-GGA calculations 
are reported for both the formation energy and the total spin moment of the pseudo-binary Heusler alloy Co$_{2-x}$Mn$_{1+x}$Si. These cluster expansions have been employed to study the  thermodynamics of this alloy and its magnetic properties at various compositions.

Both analytical calculations and Monte Carlo simulations for \cms close to the stoichiometric composition show that its thermodynamic ground state is almost perfectly ordered up to temperatures of about 1000 K, with a Co--Mn swap concentration below 1\textperthousand{} in thermodynamic equilibrium. The magnetization and especially the spin polarization at finite temperatures have been investigated with an extended Heisenberg model (mapped to exchange interactions obtained from DFT calculations) and Monte Carlo simulations. It has been argued that the spin polarization of ordered \cms is nearly 100\% for $T < 0.27~T_{\text{C}}$ (265~K). \cite{Lezaic2006} Therefore, it should be possible to prepare samples (e.g. by careful annealing) that have a high structural order (avoiding the defect states) and indeed display the theoretically predicted spin gap at low temperatures. Adding more Mn to \cms can be used to keep the Co antisite concentration low even at higher temperatures. For non-stoichiometric, moderately Mn-rich \cms we have shown that the deficiency of electrons due to replacing Co atoms by Mn atoms is fully accommodated in the majority spin channel, and thus does not affect the half-metallic gap (extended Slater-Pauling rule). Using the  cluster expansion for the total spin moment, we identify a large interval of Mn concentrations ranging from 25\% to 60\% where potentially half-metallic compositions may occur.

At low temperature, the cluster expansion for the total energy predicts four new ground state structures besides the well-known L2$_1$ structure. These new ordered structures are slightly lower in energy than a random occupation of Mn and Co sites. One of these new ground states, Co$_2$Mn$_4$Si$_2$, even shows a spin gap. However, Monte Carlo simulations confirm the expectation that the ordering energy associated with these structures is too small to render them stable at room temperature. Instead they decompose into regions of  \cms and Mn$_3$Si.

\section*{Acknowledgment}
P.K. acknowledges financial support by the Deutsche Forschungsgemeinschaft (DFG) within Research Center Sfb491.


\begin{thebibliography}{39}
\expandafter\ifx\csname natexlab\endcsname\relax\def\natexlab#1{#1}\fi
\expandafter\ifx\csname bibnamefont\endcsname\relax
  \def\bibnamefont#1{#1}\fi
\expandafter\ifx\csname bibfnamefont\endcsname\relax
  \def\bibfnamefont#1{#1}\fi
\expandafter\ifx\csname citenamefont\endcsname\relax
  \def\citenamefont#1{#1}\fi
\expandafter\ifx\csname url\endcsname\relax
  \def\url#1{\texttt{#1}}\fi
\expandafter\ifx\csname urlprefix\endcsname\relax\def\urlprefix{URL }\fi
\providecommand{\bibinfo}[2]{#2}
\providecommand{\eprint}[2][]{\url{#2}}

\bibitem[{\citenamefont{de~Groot et~al.}(1983)\citenamefont{de~Groot, Mueller,
  van Engen, and Buschow}}]{deGroot1983}
\bibinfo{author}{\bibfnamefont{R.~A.} \bibnamefont{de~Groot}},
  \bibinfo{author}{\bibfnamefont{F.~M.} \bibnamefont{Mueller}},
  \bibinfo{author}{\bibfnamefont{P.~G.} \bibnamefont{van Engen}},
  \bibnamefont{and} \bibinfo{author}{\bibfnamefont{K.~H.~J.}
  \bibnamefont{Buschow}}, \bibinfo{journal}{Phys. Rev. Lett.}
  \textbf{\bibinfo{volume}{50}}, \bibinfo{pages}{2024} (\bibinfo{year}{1983}).

\bibitem[{\citenamefont{Chioncel et~al.}(2003)\citenamefont{Chioncel,
  Katsnelson, deGroot, and Lichtenstein}}]{Chioncel2003}
\bibinfo{author}{\bibfnamefont{L.}~\bibnamefont{Chioncel}},
  \bibinfo{author}{\bibfnamefont{M.~I.} \bibnamefont{Katsnelson}},
  \bibinfo{author}{\bibfnamefont{R.~A.} \bibnamefont{deGroot}},
  \bibnamefont{and} \bibinfo{author}{\bibfnamefont{A.~I.}
  \bibnamefont{Lichtenstein}}, \bibinfo{journal}{Phys. Rev. B}
  \textbf{\bibinfo{volume}{68}}, \bibinfo{pages}{144425}
  (\bibinfo{year}{2003}).

\bibitem[{\citenamefont{Chioncel et~al.}(2006)\citenamefont{Chioncel, Arrigoni,
  Katsnelson, and Lichtenstein}}]{Chioncel2006}
\bibinfo{author}{\bibfnamefont{L.}~\bibnamefont{Chioncel}},
  \bibinfo{author}{\bibfnamefont{E.}~\bibnamefont{Arrigoni}},
  \bibinfo{author}{\bibfnamefont{M.~I.} \bibnamefont{Katsnelson}},
  \bibnamefont{and} \bibinfo{author}{\bibfnamefont{A.~I.}
  \bibnamefont{Lichtenstein}}, \bibinfo{journal}{Phys. Rev. Lett.}
  \textbf{\bibinfo{volume}{96}}, \bibinfo{pages}{137203}
  (\bibinfo{year}{2006}).

\bibitem[{\citenamefont{Ritchie et~al.}(2003)\citenamefont{Ritchie, Xiao, Ji,
  Chen, Chien, Zhang, Chen, Liu, Wu, and Zhang}}]{Ritchie2003}
\bibinfo{author}{\bibfnamefont{L.}~\bibnamefont{Ritchie}},
  \bibinfo{author}{\bibfnamefont{G.}~\bibnamefont{Xiao}},
  \bibinfo{author}{\bibfnamefont{Y.}~\bibnamefont{Ji}},
  \bibinfo{author}{\bibfnamefont{T.~Y.} \bibnamefont{Chen}},
  \bibinfo{author}{\bibfnamefont{C.~L.} \bibnamefont{Chien}},
  \bibinfo{author}{\bibfnamefont{M.}~\bibnamefont{Zhang}},
  \bibinfo{author}{\bibfnamefont{J.}~\bibnamefont{Chen}},
  \bibinfo{author}{\bibfnamefont{Z.}~\bibnamefont{Liu}},
  \bibinfo{author}{\bibfnamefont{G.}~\bibnamefont{Wu}}, \bibnamefont{and}
  \bibinfo{author}{\bibfnamefont{X.~X.} \bibnamefont{Zhang}},
  \bibinfo{journal}{Phys. Rev. B} \textbf{\bibinfo{volume}{68}},
  \bibinfo{pages}{104430} (\bibinfo{year}{2003}).

\bibitem[{\citenamefont{Singh et~al.}(2004)\citenamefont{Singh, Barber,
  Miyoshi, Bugoslavsky, Branford, and Cohen}}]{Singh2004b}
\bibinfo{author}{\bibfnamefont{L.~J.} \bibnamefont{Singh}},
  \bibinfo{author}{\bibfnamefont{Z.~H.} \bibnamefont{Barber}},
  \bibinfo{author}{\bibfnamefont{Y.}~\bibnamefont{Miyoshi}},
  \bibinfo{author}{\bibfnamefont{Y.}~\bibnamefont{Bugoslavsky}},
  \bibinfo{author}{\bibfnamefont{W.~R.} \bibnamefont{Branford}},
  \bibnamefont{and} \bibinfo{author}{\bibfnamefont{L.~F.} \bibnamefont{Cohen}},
  \bibinfo{journal}{Appl. Phys. Lett.} \textbf{\bibinfo{volume}{84}},
  \bibinfo{pages}{2367} (\bibinfo{year}{2004}).

\bibitem[{\citenamefont{K{\"a}mmerer et~al.}(2004)\citenamefont{K{\"a}mmerer,
  Thomas, H{\"u}tten, and Reiss}}]{Kaemmerer2004}
\bibinfo{author}{\bibfnamefont{S.}~\bibnamefont{K{\"a}mmerer}},
  \bibinfo{author}{\bibfnamefont{A.}~\bibnamefont{Thomas}},
  \bibinfo{author}{\bibfnamefont{A.}~\bibnamefont{H{\"u}tten}},
  \bibnamefont{and} \bibinfo{author}{\bibfnamefont{G.}~\bibnamefont{Reiss}},
  \bibinfo{journal}{Appl. Phys. Lett.} \textbf{\bibinfo{volume}{85}},
  \bibinfo{pages}{79} (\bibinfo{year}{2004}).

\bibitem[{\citenamefont{Singh et~al.}(2006)\citenamefont{Singh, Barber, Kohn,
  Petford-Long, Miyoshi, Bugoslavsky, and Cohen}}]{Singh2006}
\bibinfo{author}{\bibfnamefont{L.~J.} \bibnamefont{Singh}},
  \bibinfo{author}{\bibfnamefont{Z.~H.} \bibnamefont{Barber}},
  \bibinfo{author}{\bibfnamefont{A.}~\bibnamefont{Kohn}},
  \bibinfo{author}{\bibfnamefont{A.~K.} \bibnamefont{Petford-Long}},
  \bibinfo{author}{\bibfnamefont{Y.}~\bibnamefont{Miyoshi}},
  \bibinfo{author}{\bibfnamefont{Y.}~\bibnamefont{Bugoslavsky}},
  \bibnamefont{and} \bibinfo{author}{\bibfnamefont{L.~F.} \bibnamefont{Cohen}},
  \bibinfo{journal}{J. Appl. Phys.} \textbf{\bibinfo{volume}{99}},
  \bibinfo{pages}{013904} (\bibinfo{year}{2006}).

\bibitem[{\citenamefont{Orgassa et~al.}(1999)\citenamefont{Orgassa, Fujiwara,
  Schulthess, and Butler}}]{Orgassa1999}
\bibinfo{author}{\bibfnamefont{D.}~\bibnamefont{Orgassa}},
  \bibinfo{author}{\bibfnamefont{H.}~\bibnamefont{Fujiwara}},
  \bibinfo{author}{\bibfnamefont{T.~C.} \bibnamefont{Schulthess}},
  \bibnamefont{and} \bibinfo{author}{\bibfnamefont{W.~H.}
  \bibnamefont{Butler}}, \bibinfo{journal}{Phys. Rev. B}
  \textbf{\bibinfo{volume}{60}}, \bibinfo{pages}{13237} (\bibinfo{year}{1999}).

\bibitem[{\citenamefont{Picozzi et~al.}(2004)\citenamefont{Picozzi, Continenza,
  and Freeman}}]{Picozzi2004a}
\bibinfo{author}{\bibfnamefont{S.}~\bibnamefont{Picozzi}},
  \bibinfo{author}{\bibfnamefont{A.}~\bibnamefont{Continenza}},
  \bibnamefont{and} \bibinfo{author}{\bibfnamefont{A.~J.}
  \bibnamefont{Freeman}}, \bibinfo{journal}{Phys. Rev. B}
  \textbf{\bibinfo{volume}{69}}, \bibinfo{pages}{094423}
  (\bibinfo{year}{2004}).

\bibitem[{\citenamefont{Ishida et~al.}(1995)\citenamefont{Ishida, Fujii,
  Kashiwagi, and Asano}}]{Ishida1995b}
\bibinfo{author}{\bibfnamefont{S.}~\bibnamefont{Ishida}},
  \bibinfo{author}{\bibfnamefont{S.}~\bibnamefont{Fujii}},
  \bibinfo{author}{\bibfnamefont{S.}~\bibnamefont{Kashiwagi}},
  \bibnamefont{and} \bibinfo{author}{\bibfnamefont{S.}~\bibnamefont{Asano}},
  \bibinfo{journal}{J. Phys. Soc. Jpn.} \textbf{\bibinfo{volume}{64}},
  \bibinfo{pages}{2152} (\bibinfo{year}{1995}).

\bibitem[{\citenamefont{Picozzi et~al.}(2002)\citenamefont{Picozzi, Continenza,
  and Freeman}}]{Picozzi2002}
\bibinfo{author}{\bibfnamefont{S.}~\bibnamefont{Picozzi}},
  \bibinfo{author}{\bibfnamefont{A.}~\bibnamefont{Continenza}},
  \bibnamefont{and} \bibinfo{author}{\bibfnamefont{A.~J.}
  \bibnamefont{Freeman}}, \bibinfo{journal}{Phys. Rev. B}
  \textbf{\bibinfo{volume}{66}}, \bibinfo{pages}{094421}
  (\bibinfo{year}{2002}).

\bibitem[{\citenamefont{Webster}(1971)}]{Webster1971}
\bibinfo{author}{\bibfnamefont{P.~J.} \bibnamefont{Webster}},
  \bibinfo{journal}{J. Phys. Chem. Solids} \textbf{\bibinfo{volume}{32}},
  \bibinfo{pages}{1221} (\bibinfo{year}{1971}).

\bibitem[{\citenamefont{Brown et~al.}(2000)\citenamefont{Brown, Neumann,
  Webster, and Ziebeck}}]{Brown2000}
\bibinfo{author}{\bibfnamefont{P.~J.} \bibnamefont{Brown}},
  \bibinfo{author}{\bibfnamefont{K.~U.} \bibnamefont{Neumann}},
  \bibinfo{author}{\bibfnamefont{P.~J.} \bibnamefont{Webster}},
  \bibnamefont{and} \bibinfo{author}{\bibfnamefont{K.~R.~A.}
  \bibnamefont{Ziebeck}}, \bibinfo{journal}{J. Phys.: Condens. Matter}
  \textbf{\bibinfo{volume}{12}}, \bibinfo{pages}{1827} (\bibinfo{year}{2000}).

\bibitem[{\citenamefont{Schmalhorst et~al.}(2005)\citenamefont{Schmalhorst,
  K{\"a}mmerer, Reiss, and H{\"u}tten}}]{Schmalhorst2005}
\bibinfo{author}{\bibfnamefont{J.}~\bibnamefont{Schmalhorst}},
  \bibinfo{author}{\bibfnamefont{S.}~\bibnamefont{K{\"a}mmerer}},
  \bibinfo{author}{\bibfnamefont{G.}~\bibnamefont{Reiss}}, \bibnamefont{and}
  \bibinfo{author}{\bibfnamefont{A.}~\bibnamefont{H{\"u}tten}},
  \bibinfo{journal}{Appl. Phys. Lett.} \textbf{\bibinfo{volume}{86}},
  \bibinfo{pages}{052501} (\bibinfo{year}{2005}).

\bibitem[{\citenamefont{Sakuraba et~al.}(2005)\citenamefont{Sakuraba, Nakata,
  Oogane, Kubota, Ando, Sakuma, and Miyazaki}}]{Sakuraba2005}
\bibinfo{author}{\bibfnamefont{Y.}~\bibnamefont{Sakuraba}},
  \bibinfo{author}{\bibfnamefont{J.}~\bibnamefont{Nakata}},
  \bibinfo{author}{\bibfnamefont{M.}~\bibnamefont{Oogane}},
  \bibinfo{author}{\bibfnamefont{H.}~\bibnamefont{Kubota}},
  \bibinfo{author}{\bibfnamefont{Y.}~\bibnamefont{Ando}},
  \bibinfo{author}{\bibfnamefont{A.}~\bibnamefont{Sakuma}}, \bibnamefont{and}
  \bibinfo{author}{\bibfnamefont{T.}~\bibnamefont{Miyazaki}},
  \bibinfo{journal}{Jap. J. Appl. Phys.} \textbf{\bibinfo{volume}{44}},
  \bibinfo{pages}{L1100} (\bibinfo{year}{2005}).

\bibitem[{\citenamefont{Sakuraba et~al.}(2006)\citenamefont{Sakuraba, Hattori,
  Oogane, Ando, Kato, Sakuma, Miyazaki, and Kubota}}]{Sakuraba2006a}
\bibinfo{author}{\bibfnamefont{Y.}~\bibnamefont{Sakuraba}},
  \bibinfo{author}{\bibfnamefont{M.}~\bibnamefont{Hattori}},
  \bibinfo{author}{\bibfnamefont{M.}~\bibnamefont{Oogane}},
  \bibinfo{author}{\bibfnamefont{Y.}~\bibnamefont{Ando}},
  \bibinfo{author}{\bibfnamefont{H.}~\bibnamefont{Kato}},
  \bibinfo{author}{\bibfnamefont{A.}~\bibnamefont{Sakuma}},
  \bibinfo{author}{\bibfnamefont{T.}~\bibnamefont{Miyazaki}}, \bibnamefont{and}
  \bibinfo{author}{\bibfnamefont{H.}~\bibnamefont{Kubota}},
  \bibinfo{journal}{Appl. Phys. Lett.} \textbf{\bibinfo{volume}{88}},
  \bibinfo{pages}{192508} (\bibinfo{year}{2006}).

\bibitem[{\citenamefont{Ishikawa et~al.}(2006)\citenamefont{Ishikawa, Marukame,
  Kijima, Matsuda, Uemura, Arita, and Yamamoto}}]{Ishikawa2006}
\bibinfo{author}{\bibfnamefont{T.}~\bibnamefont{Ishikawa}},
  \bibinfo{author}{\bibfnamefont{T.}~\bibnamefont{Marukame}},
  \bibinfo{author}{\bibfnamefont{H.}~\bibnamefont{Kijima}},
  \bibinfo{author}{\bibfnamefont{K.-I.} \bibnamefont{Matsuda}},
  \bibinfo{author}{\bibfnamefont{T.}~\bibnamefont{Uemura}},
  \bibinfo{author}{\bibfnamefont{M.}~\bibnamefont{Arita}}, \bibnamefont{and}
  \bibinfo{author}{\bibfnamefont{M.}~\bibnamefont{Yamamoto}},
  \bibinfo{journal}{Appl. Phys. Lett.} \textbf{\bibinfo{volume}{89}},
  \bibinfo{pages}{192505} (\bibinfo{year}{2006}).

\bibitem[{\citenamefont{Schmalhorst et~al.}(2007)\citenamefont{Schmalhorst,
  Thomas, K{\"a}mmerer, Schebaum, Ebke, Sacher, Reiss, H{\"u}tten, Turchanin,
  G{\"o}lzh{\"a}user, and Arenholz}}]{Schmalhorst2007}
\bibinfo{author}{\bibfnamefont{J.}~\bibnamefont{Schmalhorst}},
  \bibinfo{author}{\bibfnamefont{A.}~\bibnamefont{Thomas}},
  \bibinfo{author}{\bibfnamefont{S.}~\bibnamefont{K{\"a}mmerer}},
  \bibinfo{author}{\bibfnamefont{O.}~\bibnamefont{Schebaum}},
  \bibinfo{author}{\bibfnamefont{D.}~\bibnamefont{Ebke}},
  \bibinfo{author}{\bibfnamefont{M.~D.} \bibnamefont{Sacher}},
  \bibinfo{author}{\bibfnamefont{G.}~\bibnamefont{Reiss}},
  \bibinfo{author}{\bibfnamefont{A.}~\bibnamefont{H{\"u}tten}},
  \bibinfo{author}{\bibfnamefont{A.}~\bibnamefont{Turchanin}},
  \bibinfo{author}{\bibfnamefont{A.}~\bibnamefont{G{\"o}lzh{\"a}user}},  \bibnamefont{and} 
  \bibinfo{author}{\bibfnamefont{E.}~\bibnamefont{Arenholz}}, 
  \bibinfo{journal}{Phys. Rev. B}
  \textbf{\bibinfo{volume}{75}}, \bibinfo{pages}{014403}
  (\bibinfo{year}{2007}).

\bibitem[{\citenamefont{Tsunegi et~al.}(2006)\citenamefont{Tsunegi, Sakuraba, Hattori, Oogane, Takanashi, and Ando}}]{Tsunegi2008}
  \bibinfo{author}{\bibfnamefont{S.}~\bibnamefont{Tsunegi}},
  \bibinfo{author}{\bibfnamefont{Y.}~\bibnamefont{Sakuraba}},
  \bibinfo{author}{\bibfnamefont{M.}~\bibnamefont{Oogane}},
    \bibinfo{author}{\bibfnamefont{K.}~\bibnamefont{Takanashi}},
    \bibnamefont{and}
  \bibinfo{author}{\bibfnamefont{Y.}~\bibnamefont{Ando}},
  \bibinfo{journal}{Appl. Phys. Lett.} \textbf{\bibinfo{volume}{93}},
  \bibinfo{pages}{112506} (\bibinfo{year}{2008}).

\bibitem[{\citenamefont{Ravel et~al.}(2002)\citenamefont{Ravel, Raphael,
  Harris, and Huang}}]{Ravel2002a}
\bibinfo{author}{\bibfnamefont{B.}~\bibnamefont{Ravel}},
  \bibinfo{author}{\bibfnamefont{M.~P.} \bibnamefont{Raphael}},
  \bibinfo{author}{\bibfnamefont{V.~G.} \bibnamefont{Harris}},
  \bibnamefont{and} \bibinfo{author}{\bibfnamefont{Q.}~\bibnamefont{Huang}},
  \bibinfo{journal}{Phys. Rev. B} \textbf{\bibinfo{volume}{65}},
  \bibinfo{pages}{184431} (\bibinfo{year}{2002}).

\bibitem[{\citenamefont{Picozzi and Freeman}(2007)}]{Picozzi2007}
\bibinfo{author}{\bibfnamefont{S.}~\bibnamefont{Picozzi}} \bibnamefont{and}
  \bibinfo{author}{\bibfnamefont{A.~J.} \bibnamefont{Freeman}},
  \bibinfo{journal}{J. Phys.: Condens. Matter} \textbf{\bibinfo{volume}{19}},
  \bibinfo{pages}{315215} (\bibinfo{year}{2007}).

\bibitem[{\citenamefont{Sanchez et~al.}(1984)\citenamefont{Sanchez, Ducastelle,
  and Gratias}}]{Sanchez1984}
\bibinfo{author}{\bibfnamefont{J.~M.} \bibnamefont{Sanchez}},
  \bibinfo{author}{\bibfnamefont{F.}~\bibnamefont{Ducastelle}},
  \bibnamefont{and} \bibinfo{author}{\bibfnamefont{D.}~\bibnamefont{Gratias}},
  \bibinfo{journal}{Physica A} \textbf{\bibinfo{volume}{128}},
  \bibinfo{pages}{334} (\bibinfo{year}{1984}).

\bibitem[{\citenamefont{M{\"u}ller}(2003)}]{Mueller2003}
\bibinfo{author}{\bibfnamefont{S.}~\bibnamefont{M{\"u}ller}},
  \bibinfo{journal}{J. Phys.: Condens. Matter} \textbf{\bibinfo{volume}{15}},
  \bibinfo{pages}{R1429} (\bibinfo{year}{2003}).

\bibitem[{\citenamefont{van~de Walle and Ceder}(2002)}]{vandeWalle2002a}
\bibinfo{author}{\bibfnamefont{A.}~\bibnamefont{van~de Walle}}
  \bibnamefont{and} \bibinfo{author}{\bibfnamefont{G.}~\bibnamefont{Ceder}},
  \bibinfo{journal}{J. Phase Equilibria Diffus.} \textbf{\bibinfo{volume}{23}},
  \bibinfo{pages}{348} (\bibinfo{year}{2002}).

\bibitem[{\citenamefont{Connolly and Williams}(1983)}]{Connolly1983}
\bibinfo{author}{\bibfnamefont{J.~W.~D.} \bibnamefont{Connolly}}
  \bibnamefont{and} \bibinfo{author}{\bibfnamefont{A.~R.}
  \bibnamefont{Williams}}, \bibinfo{journal}{Phys. Rev. B}
  \textbf{\bibinfo{volume}{27}}, \bibinfo{pages}{5169} (\bibinfo{year}{1983}).

\bibitem[{\citenamefont{Ozoli{\c n}{\u s} et~al.}(1998)\citenamefont{Ozoli{\c
  n}{\u s}, Wolverton, and Zunger}}]{Ozolins1998}
\bibinfo{author}{\bibfnamefont{V.}~\bibnamefont{Ozoli{\c n}{\u s}}},
  \bibinfo{author}{\bibfnamefont{C.}~\bibnamefont{Wolverton}},
  \bibnamefont{and} \bibinfo{author}{\bibfnamefont{A.}~\bibnamefont{Zunger}},
  \bibinfo{journal}{Phys. Rev. B} \textbf{\bibinfo{volume}{57}},
  \bibinfo{pages}{6427} (\bibinfo{year}{1998}).

\bibitem[{\citenamefont{Van~der Ven and Ceder}(2005)}]{vanderVen2005}
\bibinfo{author}{\bibfnamefont{A.}~\bibnamefont{Van~der Ven}} \bibnamefont{and}
  \bibinfo{author}{\bibfnamefont{G.}~\bibnamefont{Ceder}},
  \bibinfo{journal}{Phys. Rev. B} \textbf{\bibinfo{volume}{71}},
  \bibinfo{pages}{054102} (\bibinfo{year}{2005}).

\bibitem[{\citenamefont{D\'iaz-Ortiz et~al.}(2006)\citenamefont{D\'iaz-Ortiz,
  Drautz, F{\"a}hnle, Dosch, and Sanchez}}]{Ortiz2006}
\bibinfo{author}{\bibfnamefont{A.}~\bibnamefont{D\'iaz-Ortiz}},
  \bibinfo{author}{\bibfnamefont{R.}~\bibnamefont{Drautz}},
  \bibinfo{author}{\bibfnamefont{M.}~\bibnamefont{F{\"a}hnle}},
  \bibinfo{author}{\bibfnamefont{H.}~\bibnamefont{Dosch}}, \bibnamefont{and}
  \bibinfo{author}{\bibfnamefont{J.~M.} \bibnamefont{Sanchez}},
  \bibinfo{journal}{Phys. Rev. B} \textbf{\bibinfo{volume}{73}},
  \bibinfo{pages}{224208} (\bibinfo{year}{2006}).

\bibitem[{\citenamefont{Franceschetti et~al.}(2006)\citenamefont{Franceschetti,
  Dudiy, Barabash, Zunger, Xu, and van Schilfgaarde}}]{Franceschetti2006}
\bibinfo{author}{\bibfnamefont{A.}~\bibnamefont{Franceschetti}},
  \bibinfo{author}{\bibfnamefont{S.~V.} \bibnamefont{Dudiy}},
  \bibinfo{author}{\bibfnamefont{S.~V.} \bibnamefont{Barabash}},
  \bibinfo{author}{\bibfnamefont{A.}~\bibnamefont{Zunger}},
  \bibinfo{author}{\bibfnamefont{J.}~\bibnamefont{Xu}}, \bibnamefont{and}
  \bibinfo{author}{\bibfnamefont{M.}~\bibnamefont{van Schilfgaarde}},
  \bibinfo{journal}{Phys. Rev. Lett.} \textbf{\bibinfo{volume}{97}},
  \bibinfo{pages}{047202} (\bibinfo{year}{2006}).

\bibitem[{\citenamefont{Schwarz and Blaha}(2003)}]{Schwarz2003}
\bibinfo{author}{\bibfnamefont{K.}~\bibnamefont{Schwarz}} \bibnamefont{and}
  \bibinfo{author}{\bibfnamefont{P.}~\bibnamefont{Blaha}},
  \bibinfo{journal}{Comput. Mater. Sci.} \textbf{\bibinfo{volume}{28}},
  \bibinfo{pages}{259} (\bibinfo{year}{2003}).

\bibitem[{\citenamefont{Perdew et~al.}(1996)\citenamefont{Perdew, Burke, and
  Ernzerhof}}]{Perdew1996}
\bibinfo{author}{\bibfnamefont{J.~P.} \bibnamefont{Perdew}},
  \bibinfo{author}{\bibfnamefont{K.}~\bibnamefont{Burke}}, \bibnamefont{and}
  \bibinfo{author}{\bibfnamefont{M.}~\bibnamefont{Ernzerhof}},
  \bibinfo{journal}{Phys. Rev. Lett.} \textbf{\bibinfo{volume}{77}},
  \bibinfo{pages}{3865} (\bibinfo{year}{1996}).

\bibitem[{\citenamefont{Hashemifar et~al.}(2005)\citenamefont{Hashemifar,
  Kratzer, and Scheffler}}]{Hashemifar2005}
\bibinfo{author}{\bibfnamefont{S.~J.} \bibnamefont{Hashemifar}},
  \bibinfo{author}{\bibfnamefont{P.}~\bibnamefont{Kratzer}}, \bibnamefont{and}
  \bibinfo{author}{\bibfnamefont{M.}~\bibnamefont{Scheffler}},
  \bibinfo{journal}{Phys. Rev. Lett.} \textbf{\bibinfo{volume}{94}},
  \bibinfo{pages}{096402} (\bibinfo{year}{2005}).

\bibitem[{\citenamefont{Galanakis et~al.}(2002)\citenamefont{Galanakis,
  Dederichs, and Papanikolaou}}]{Galanakis2002b}
\bibinfo{author}{\bibfnamefont{I.}~\bibnamefont{Galanakis}},
  \bibinfo{author}{\bibfnamefont{P.~H.} \bibnamefont{Dederichs}},
  \bibnamefont{and}
  \bibinfo{author}{\bibfnamefont{N.}~\bibnamefont{Papanikolaou}},
  \bibinfo{journal}{Phys. Rev. B} \textbf{\bibinfo{volume}{66}},
  \bibinfo{pages}{174429} (\bibinfo{year}{2002}).

\bibitem[{\citenamefont{van~de Walle and Neugebauer}(2004)}]{CvandeWalle2004}
\bibinfo{author}{\bibfnamefont{C.~G.} \bibnamefont{van~de Walle}}
  \bibnamefont{and}
  \bibinfo{author}{\bibfnamefont{J.}~\bibnamefont{Neugebauer}},
  \bibinfo{journal}{J. Appl. Phys.} \textbf{\bibinfo{volume}{95}},
  \bibinfo{pages}{3851} (\bibinfo{year}{2004}).

\bibitem[{\citenamefont{Hobbs and Hafner}(2001)}]{Hobbs2001}
\bibinfo{author}{\bibfnamefont{D.}~\bibnamefont{Hobbs}} \bibnamefont{and}
  \bibinfo{author}{\bibfnamefont{J.}~\bibnamefont{Hafner}},
  \bibinfo{journal}{J. Phys.: Condens. Matter} \textbf{\bibinfo{volume}{13}},
  \bibinfo{pages}{L681} (\bibinfo{year}{2001}).
  
\bibitem[{\citenamefont{
  {\"O}zdo{\~g}an, {\c S}a{\c s}io{\~g}lu}}, Galanakis, and Akta{\c s}]{Ozdogan2007a}
  \bibinfo{author}{\bibfnamefont{K.}~\bibnamefont{{\"O}zdo{\~g}an}},
    \bibinfo{author}{\bibfnamefont{I.}~\bibnamefont{Galanakis}},
 \bibinfo{author}{\bibfnamefont{E.}~\bibnamefont{{\c S}a{\c
  s}io{\~g}lu}}, 
    \bibnamefont{and}
  \bibinfo{author}{\bibfnamefont{B.}~\bibnamefont{Akta{\c s}}},
  \bibinfo{journal}{Solid State Commun.}
  \textbf{\bibinfo{volume}{142}}, \bibinfo{pages}{492}
  (\bibinfo{year}{2007}).


\bibitem[{\citenamefont{Kandpal et~al.}(2006)\citenamefont{Kandpal, Fecher,
  Felser, and Sch{\"o}nhense}}]{Kandpal2006a}
\bibinfo{author}{\bibfnamefont{H.~C.} \bibnamefont{Kandpal}},
  \bibinfo{author}{\bibfnamefont{G.~H.} \bibnamefont{Fecher}},
  \bibinfo{author}{\bibfnamefont{C.}~\bibnamefont{Felser}}, \bibnamefont{and}
  \bibinfo{author}{\bibfnamefont{G.}~\bibnamefont{Sch{\"o}nhense}},
  \bibinfo{journal}{Phys. Rev. B} \textbf{\bibinfo{volume}{73}},
  \bibinfo{pages}{094422} (\bibinfo{year}{2006}).

\bibitem[{\citenamefont{Galanakis et~al.}(2006)\citenamefont{Galanakis,
  {\"O}zdo{\~g}an, Akta{\c s}, and {\c S}a{\c s}io{\~g}lu}}]{Galanakis2006a}
\bibinfo{author}{\bibfnamefont{I.}~\bibnamefont{Galanakis}},
  \bibinfo{author}{\bibfnamefont{K.}~\bibnamefont{{\"O}zdo{\~g}an}},
  \bibinfo{author}{\bibfnamefont{B.}~\bibnamefont{Akta{\c s}}},
  \bibnamefont{and} \bibinfo{author}{\bibfnamefont{E.}~\bibnamefont{{\c S}a{\c
  s}io{\~g}lu}}, \bibinfo{journal}{Appl. Phys. Lett.}
  \textbf{\bibinfo{volume}{89}}, \bibinfo{pages}{042502}
  (\bibinfo{year}{2006}).

\bibitem[{\citenamefont{Galanakis et~al.}(2006)\citenamefont{
  {\"O}zdo{\~g}an, {\c S}a{\c s}io{\~g}lu}}, and Galanakis]{Ozdogan2007}
  \bibinfo{author}{\bibfnamefont{K.}~\bibnamefont{{\"O}zdo{\~g}an}},
 \bibinfo{author}{\bibfnamefont{E.}~\bibnamefont{{\c S}a{\c
  s}io{\~g}lu}}, 
    \bibnamefont{and}
  \bibinfo{author}{\bibfnamefont{I.}~\bibnamefont{Galanakis}},
  \bibinfo{journal}{phys. stat. sol. (RRL)}
  \textbf{\bibinfo{volume}{1}}, \bibinfo{pages}{184}
  (\bibinfo{year}{2007}).


\bibitem[{\citenamefont{van~de Walle and Asta}(2002)}]{vandeWalle2002b}
\bibinfo{author}{\bibfnamefont{A.}~\bibnamefont{van~de Walle}}
  \bibnamefont{and} \bibinfo{author}{\bibfnamefont{M.}~\bibnamefont{Asta}},
  \bibinfo{journal}{Modell. Simul. Mater. Sci. Eng.} \textbf{\bibinfo{volume}{10}},
  \bibinfo{pages}{521} (\bibinfo{year}{2002}).

\bibitem[{\citenamefont{van~de Walle et~al.}(2002)\citenamefont{van~de Walle,
  Asta, and Ceder}}]{vandeWalle2002c}
\bibinfo{author}{\bibfnamefont{A.}~\bibnamefont{van~de Walle}},
  \bibinfo{author}{\bibfnamefont{M.}~\bibnamefont{Asta}}, \bibnamefont{and}
  \bibinfo{author}{\bibfnamefont{G.}~\bibnamefont{Ceder}},
  \bibinfo{journal}{CALPHAD: Comput. Coupling Phase Diagrams Thermochem.} \textbf{\bibinfo{volume}{26}},
  \bibinfo{pages}{539} (\bibinfo{year}{2002}).

\bibitem[{\citenamefont{Le{\v z}ai{\'c} et~al.}(2006)\citenamefont{Le{\v
  z}ai{\'c}, Mavropoulos, Enkovaara, Bihlmayer, and Bl{\"u}gel}}]{Lezaic2006}
\bibinfo{author}{\bibfnamefont{M.}~\bibnamefont{Le{\v z}ai{\'c}}},
  \bibinfo{author}{\bibfnamefont{P.}~\bibnamefont{Mavropoulos}},
  \bibinfo{author}{\bibfnamefont{J.}~\bibnamefont{Enkovaara}},
  \bibinfo{author}{\bibfnamefont{G.}~\bibnamefont{Bihlmayer}},
  \bibnamefont{and}
  \bibinfo{author}{\bibfnamefont{S.}~\bibnamefont{Bl{\"u}gel}},
  \bibinfo{journal}{Phys. Rev. Lett.} \textbf{\bibinfo{volume}{97}},
  \bibinfo{pages}{026404} (\bibinfo{year}{2006}).

\end{thebibliography}
\end{document}